\documentclass[12pt]{amsart}
\usepackage{amssymb}
\usepackage{amsmath,amsthm,amsfonts}
\usepackage[applemac]{inputenc}
\usepackage{graphicx}
\usepackage{bbm}
\usepackage{pdfsync}
\usepackage{fancyhdr}
\usepackage{mathrsfs}
\usepackage{slashed}
\usepackage{color}
\usepackage{xcolor}
\usepackage[breaklinks,colorlinks,backref]{hyperref}

\hypersetup{
colorlinks, 
linktoc=all, 
linkcolor=blue, 
citecolor=hyptxt,
urlcolor=blue}
\hypersetup{
citebordercolor=,
filebordercolor=green,
linkbordercolor=blue
}
\definecolor{hyptxt}{rgb}{0.7, 0.4, 0.9}

\usepackage{bibtopic}

\usepackage[full]{textcomp} 
\usepackage{fbb} 
\usepackage[varqu,varl]{inconsolata}
\usepackage[libertine,bigdelims,vvarbb]{newtxmath}
\usepackage[cal=boondoxo]{mathalfa}
\usepackage[T1]{fontenc}
\useosf
\newcommand{\red}[1]{\textcolor{red}{#1}}

\newtheorem{prop}{Proposition}[section]

\newcommand{\beprop}{\begin{prop}}
\newcommand{\enprop}{\end{prop}}
\newcommand{\bprf}{\begin{proof}}
\newcommand{\eprf}{\end{proof}}

\newcommand{\ket}[1]{|\kern.3ex#1\kern.3ex\rangle}
\newcommand{\bra}[1]{\langle\kern.3ex #1 \kern.3ex|}
\newcommand{\scalar}[2]{\langle\kern.3ex #1 \kern.3ex|\kern.3ex#2\kern.3ex\rangle}

\def\bu{\mathbbm{1}}

\def\R{\mathbb{R}}
\def\N{\mathbb{N}}
\def\C{\mathbb{C}}
\def\Z{\mathbb{Z}}

\def\lg{\langle }
\def\rg{\rangle }

\def\ii{\mathrm{i}}
\def\ud{\mathrm{d}}

\def\bu{\mathbbm{1}}

\def\ud{\mathrm{d}}
\definecolor{hervecolor}{rgb}{0.8,0,0.7}

\numberwithin{equation}{section}

\def\R{{\rm I\hspace{-.15em}R}}

\def\1{\mbox{I\hspace{-.15em}1}}

\def\b{\begin{equation}}
\def\e{\end{equation}}

\begin{document}
\date{\today}
\title[de Sitter Asymptotic states and $S$-matrix operator]{Asymptotic states and $S$-matrix operator \\ in de Sitter ambient space formalism}
\author{M.V. Takook, J.P. Gazeau, E. Huguet}

\address{\emph{ APC, UMR 7164}\\
\emph{Universit\'e Paris Cit\'e, CNRS, Astroparticule et Cosmologie,} \\
\emph{F-75013 Paris, France}}
\email{takook@apc.in2p3.fr; gazeau@apc.in2p3.fr; huguet@apc.in2p3.fr }

{\abstract{Within the de Sitter ambient space framework, the two different bases of the one-particle Hilbert space of the de Sitter group algebra are presented for the scalar case. Using field operator algebra and its Fock space construction in this formalism, we discuss the existence of asymptotic states in de Sitter QFT under an extension of the adiabatic hypothesis and prove the Fock space completeness theorem for the massive scalar field. We define the quantum state in the limit of future and past infinity on the Sitter hyperboloid in an observer-independent way. These results allow us to examine the existence of the $S$-matrix operator for de Sitter QFT in ambient space formalism, a question usually obscure in spacetime with a cosmological event horizon for a specific observer. Some similarities and differences between QFT in Minkowski and de Sitter spaces are discussed.}}

\maketitle

\vspace{0.5cm}

\tableofcontents


\section{Introduction}
\label{intr}

The classical de Sitter (dS) spacetime can be considered an acceptable approximation for modeling our universe at its evolution's late and possibly early (inflationary) stages. However, quantum theory on the dS background faces mathematical difficulties. These are mainly due to the lack of a global timelike Killing vector field. This forbids global definitions of time evolution, energy, vacuum state, and {\it etc}. Also, in dS spacetime, an observer has a cosmological event horizon. This event horizon obscures the definition of asymptotic states and, consequently, the meaning of the $S$-matrix operator for a specific observer. This article aims to revisit these questions of asymptotic states and the $S$-matrix operator in an observer-independent manner, {\it i.e.} the dS ambient space formalism. In particular, we prove the existence of asymptotic states and Fock space completeness. We examine the quantum state at the limit of future and past infinity. We then consider the $S$-matrix operator definition from an observer-independent perspective.

It is instructive to review basic facts concerning flat spacetimes with Galilean or Poincar\'e symmetries. The time translation operator is one of the generators of these groups, and it can be considered as the global timelike Killing vector field. From Noether's theorem, the corresponding constant of motion is identified with the free physical system's energy or Hamiltonian $\mathsf{H}_f$. By dealing with the quantized versions of the Galilean free particles or relativistic free field theory, one uses this Killing vector field to define the time evolution of the quantum state of the free physical system. For a time-dependent interaction case, $\mathsf{H}=\mathsf{H}_f+\mathsf{H}_i(t)$, the time translation is no longer a symmetry property of the physical system, and its energy is not a constant of motion. This situation is similar to the free physical system in dS spacetime, where there is a permanent interaction with the classical gravitational background.

In Minkowskian QFT, the existence of a global timelike Killing vector field allows us to define the asymptotic states at $t\longrightarrow \pm \infty$, the so-called in and out states, for which the interaction vanishes. Time translation symmetry is restored at this limit. The Poincar\'e invariant $S$-matrix operator is explicitly defined as $S=U(\infty,-\infty)$, where $U(t,t')$ is the time evolution operator. Moreover, the existence of the $S$-matrix depends on the existence of asymptotic states and the completeness of Hilbert space, {\it i.e.} $\mathcal{H}_{in}\equiv \mathcal{H}\equiv \mathcal{H}_{out}$. For a comprehensive review of this subject in Minkowskian QFT, see the illuminating \cite{bailse}.

In dS QFT, the lack of a global timelike Killing vector field leads to a much more elaborate situation. Nevertheless, owing to the maximally symmetric nature of this spacetime with the symmetry group SO$_0(1,4)$, many of the dS QFT problems can be solved. The use of both complex geometry, developed by Bros and his collaborators \cite{brgamo,brmo96}, and group theory, using elementary systems in the Wigner sense (see \cite{engapewa} for a review), have proved to be powerful tools for studying dS quantum field theory in the framework of dS ambient space formalism, see \cite{ta1}.

More precisely, the analytical approach in dS ambient space formalism has been an essential tool for extending crucial mathematical properties of Minkowskian QFT to dS space-time \cite{brepmo}, namely: the Reeh-Schlieder theorem, Bisognano-Wichmann theorem, the Kall\'en-Lehmann type representation \cite{brmo96,brepmo}, the Pancherel formulae on the dS hyperboloid \cite{tak,mol,mol2}, and then the Bros-Fourier-Helgason transformation \cite{brmo2}. The group theoretical approach has been, in particular, of central importance in clarifying the notion of the ``mass'' parameter ($\nu$) in dS spacetime \cite{ga}. This parameter, which labels the unitary irreducible representations (UIR) of the dS group, plays a crucial role in proving the completeness of asymptotic state spaces.

The paper is organized as follows. In Section \ref{Secmaen}, known results, mandatory for our purpose, are reviewed in the ambient space formalism. This includes the mass parameter, which is a key point in our discussion, the construction of one-particle Hilbert space, and Fock space. For simplicity, only the scalar case is considered. Using the Plancherel formulae, the Reeh-Schlieder theorem, and the Kall\'en-Lehmann type representation, we prove the asymptotic states completeness theorem in Section \ref{asycop}. The questions of the quantum state in the limit of future and past infinity and the $S$-matrix operator are addressed in Section \ref{timevop}. We conclude with some remarks in Section \ref{conclu}. In the appendix \ref{notations}, we present our notations. An orthonormal basis of $L^2(X)$ with $X$ equipped with a Klein-Gordon inner product is presented in appendix \ref{L2X}. We briefly recall in appendix \ref{ap}, Reeh-Schlieder theorem, Plancherel formulae, and Kall\'en-Lehmann type representation, in Minkowskian QFT. The Bros-Fourier-Helgason transformations for massive and massless fields are briefly described in appendices \ref{apc} and \ref{apd} respectively.


\section{Important notions} \label{Secmaen}

\subsection{Mass and rest energy parameters} \label{maen}

In Minkowski space-time, the rest energy $E^{\mathrm{rest}}$ of an elementary system is exactly its proper mass $m$ $(c=1)$. These two notions decouple in dS space-time \cite{engapewa,ga,coga21}. In addition, the notion of rest energy $E_{\mathrm{dS}}^{\mathrm{rest}}$ in dS space is ill-defined due to the ambiguity in the time definition. The proper mass $m_{\mathrm{dS}}$ in dS spacetime should be independent of spacetime geometry, or equivalently there should not exist any difference between inertial and gravitational mass in classical theory, which is a manifestation of the equivalence principle. Hence, one expect $m_{\mathrm{dS}}=m$.

At the quantum level, a massive elementary system, in the Wigner sense, in dS spacetime is described by a UIR of the dS group, $U^{(\nu,s)}$, in the so-called principal series. These UIRs are labeled by two parameters: $s\in \Z/2$ corresponding to the spin and $\nu\in \R$ which is a parameter related to a mass and to the ``rest energy'' as defined below. The parameter $\nu$ is usually employed for representations of the principal and complementary series of the dS group. $n$ is used for discrete series representations. For the principal series of the dS group, one here adopts the definition of mass in dS spacetime proposed by Garidi \cite{ga}, precisely in terms of $\nu$, $c$, the dS curvature radius $R$, and the fundamental constants $\hbar, c$:
\begin{equation}
\label{garidi}
m=\frac{\hbar}{Rc}\left[\nu^2 + \left(s-\frac{1}{2}\right)^2\right]\, .
\end{equation}
By inverting this equation, we  define the dS  ``rest energy'' in terms of  $\nu$,   $m$ and  $s$ \cite{coga21}:
\begin{equation}
\label{mEdS}
E_{\mathrm{dS}}^{\mathrm{rest}}:=\frac{ \hbar c  \nu}{R} = \pm\left[m^2c^4 -\frac{ \hbar^2 c^2 }{R^2}\left(s - \frac{1}{2}\right)^2\right]^{1/2},
\end{equation}
which, as expected, imposes
\begin{equation}
\label{lowlimm}
m \geq \frac{\hbar }{Rc} \left\vert s-\frac{1}{2}\right\vert\equiv m_R\left\vert s-\frac{1}{2}\right\vert\, .  
\end{equation}
 From the point of view of a Minkowskian tangent observer, the sign $+$ is naturally preferred. These representations are said to be ''massive'' because at the null curvature limit ($R^{-1}=0$), the rest energy merges with the proper mass, as expected. Hence, due to the curvature, for a free-falling elementary particle residing on a dS geodesic, its rest energy comes from its proper mass and spin, the latter interacting with the curvature. Note that for fermions with $s=1/2$ the dS rest energy is the Minkowskian $mc^2$, while it is equal to $mc^2\sqrt{1 - m^2_R/4m^2}$ for massive scalars and spin $1$ bosons, which imposes the lowest limit $m\geq m_R/2$ for the mass.

Definitions \eqref{garidi} and \eqref{mEdS} are consistent with the massless case but differ from the definition(s) of mass $m$ given in other works, like \cite{brepmo2}. Indeed, Eq. \eqref{garidi} yields $m=0$ for the value $\nu=\pm\ii\vert s- 1/2\vert$ assumed by the parameter $\nu$ when one deals with the massless dS UIRs with arbitrary spin (in that case helicity), namely those UIRs that have unambiguous Minkowskian counterparts ($R^{-1}=0$). The case $\nu=\pm\ii/2$, $s=0$ stands for the massless conformally coupled (mcc) case and lies in the complementary series. The latter corresponds to the values $\nu\in \ii\R$ with $0< \vert\nu\vert <3/2$ (for $s=0$) and $0< \vert\nu\vert <1/2$ (for $s\neq 0$). Therefore, in the case of the complementary series Eq. \eqref{garidi} is meaningful (no tachyons!) for $0< \vert\nu\vert <1/2$. The other massless cases for $s>0$ correspond to the representations $\Pi^{\pm}_{s s}$ (in Dixmier'notation) lying at the bottom of the discrete series, where $\pm$ stands for helicity. For the arbitrary elements of the discrete series, the scalar case, $\Pi^{\pm}_{p,0}$, with $p=1,2, \cdots$, and the other cases, $\Pi^{\pm}_{p,q}$, with $p=\frac{1}{2}, 1,\frac{3}{2}, 2, \cdots$, $q=p,p-1, \cdots, 1$ or $\frac{1}{2}$, the Garidi mass reads $m=\frac{\hbar}{Rc}[(p-q)(p+q+1)]^{1/2}$ whereas \eqref{mEdS} has no meaning for these UIRs of the dS group, particularly for the ``massless minimally coupled'' (mmc) scalar field representation $\Pi^{\pm}_{p,0}$ for which $\nu = \ii 3/2$. All details can be found in \cite{engapewa}.

To end this discussion about mass and rest energy in dS space, it is interesting to give the estimate of the limit mass $m_R$ in the function of the currently observed cosmological constant $R= \sqrt{3/\Lambda}$:
\begin{equation}
\label{estmR}
\Lambda \approx (2.1 \pm 0.1) \times 10^{-52}\, \mathrm{m}^{-2} \Rightarrow m_R = \frac{\hbar}{Rc}\approx 0.55 \times 10^{-59}\,\mathrm{kg}\approx 3.1\times 10^{-24}\,\mathrm{eV}\,.
\end{equation}
Note that this value fits the current estimation of the minimal potential mass of dark matter ``particles'' \cite{DMP}. In the sequel, we will adopt the atomic units $\hbar=1 =c$ and the Hubble notation $H= c/R=1/R$ and replace the discrete series parameter $p$ with $n$.

In Minkowskian QFT, the free one-particle scalar states, $\vert k^\mu;m\rangle$, with various proper masses $m$ ($k_\mu k^\mu=m^2$), are independent of each other, and for a specific mass $m$, they form a complete basis for the free one-particle Hilbert space. Interaction entangles quantum states with different mass parameters, so a specific mass $m$ is no longer a complete basis. An integral over different mass parameters is needed to obtain a complete basis and to build the two-point function for the interacting case, which can be seen in the Kall\'en-Lehmann spectral representation; see appendix \ref{ap}.

In dS spacetime, where there is a permanent interaction with the classical gravitational background, similar to the Minkowskian interaction field, the quantum states with a specific representation parameter $\nu$ are not sufficient to obtain a complete basis for the one-particle Hilbert space. In \cite{tak},  Takahashi gave Plancherel's formula for the dS group as a sum and an integral over $n$ and $\nu$ respectively (see also Vilenkin in \cite{vilenkin}). This formula was used by Molchanov \cite{mol,mol2} to establish, in the scalar case, an expansion of a Dirac delta distribution (see Eqs. \eqref{invdelt} -- \eqref{invdelt} for more details) on the unit one-sheeted  $d-1$-dimensional hyperboloid $X= \{x\in \R^d\, , \, -x_0^2 + x_1^2  + \cdots x_{d-1}^2 =1\}$ as:
\begin{equation}
\label{deltamol}
\delta_P (x)=\sum _{n=1}^{\infty } a_n \overline{\Psi_n(x)}+\int_{-\infty}^\infty \ud \nu\, \varpi(\nu)\,\overline{\Phi_\nu(x)} \,, 
\end{equation}
where $\delta_P (x)$ stands for the Dirac distribution with support for the equatorial point \begin{equation}
\label{eqpt}
P=(0,1,0,\cdots 0)\in X\,.
\end{equation} 
The functions $\Psi_n$ and $\Phi_\nu$ are ``spherical'' functions, which are expressed in terms of Legendre functions. 
In Eq. \eqref{deltamol}, the sum is over the discrete series parameter $n$, and the integral is over the principal series parameter $\nu$. Coefficients and integral weights are given by:
\begin{equation}
\label{coeffan}
a_n=  \frac{1}{2} \pi^{-d/2-1}\Gamma(2n+d-2)\Gamma(n+d-2) 4^{2-d-n}
\end{equation} 
\begin{equation}
\label{coeffom}
\varpi(\nu)= \frac{1}{2^{2d-3}\pi^{d-1}}\left\lbrace\begin{array}{cc}
    \nu\tanh\frac{\pi\nu}{2}\, \prod_{k=1}^{\frac{d-3}{2}} [(2k-1)^2 +\nu^2]\, ,  &  d \ \mbox{odd}  \\
 &   \\
     \prod_{k=1}^{\left\lfloor\frac{d-3}{2}\right\rfloor} (4k^2 +\nu^2)\, ,  &  d \ \mbox{even}\,.     
\end{array}\right.
\end{equation}
In the present study $d=5$ and so 
\begin{equation}
\label{coeffomd5}
a_n= \frac{1}{2\pi^{7/2}}\Gamma(2n+3)\Gamma(n+3) 4^{-3-n}\, , \quad \varpi(\nu)=\frac{1}{2^7\pi^4} \nu\tanh\frac{\pi\nu}{2}(\nu^2+1)(\nu^2+9)\, . 
\end{equation}}

Bros et al established an analogous formula for ambient space formalism in \cite{brmo96,brepmo,brmo2} through Cauchy's integral on the holomorphic functions involved in the considered representations. The use of holomorphic functions and analytical properties allows them to define the Fourier-Helgason transformation, which in the sequel, is called the Bros-Fourier-Helgason transformation, see appendix \ref{apc0}.


\subsection{One-particle states} \label{A}

As we mention in the introduction, we will restrict our analysis to the scalar fields since introducing fields with spin requires extra technicalities that are not fundamental to our purpose. Nevertheless, we note that the scalar case can be generalized to the various massive or massless fields with nonzero spin, corresponding to various (unitary irreducible) representations of the dS group.

The Hilbert space $\mathcal{H}^{(1)}$ relevant to the first quantization formalism in dS spacetime can be constructed from the dS group algebra. This construction has been considered by Thomas \cite{tho,new} and was completed by Dixmier \cite{dix}:
\b \label{algebra} [J_a,J_b]=f_{ab}^c J_c \;\; \Longrightarrow \;\; |\alpha; j_1,j_2; m_{j_1},m_{j_2}\rangle \in \mathcal{H}^{(1)}\equiv \bigoplus_{\alpha; j_1 ,j_2} \; \mathcal{H}^{\alpha; j_1,j_2}\, ,\e
where the $J_a$  are the generators of the de~Sitter group, $f_{ab}^c$ the structure constants,  $j_1$ and $j_2$ are two numbers, labeling the UIR's of the maximal compact subgroup SO($4$), picked in the sequence $0,\frac{1}{2},1,\cdots$, such that $-j\leq m_j \leq j$, and the $\alpha$'s are sets of parameters numbering the columns and rows of the (generalized) matrices, assuming continuous or discrete values \cite{tho}. For admissible values of $\alpha, j_1$ and $j_2$ giving rise to the closure of the Hilbert space $\mathcal{H}^{(1)}$ under the action of the dS group generators see \cite{tho,dix}.

Now, the one-particle Hilbert space for a spinless massive system ($\alpha\equiv\nu, j_1=j_2=0$) is defined as the direct  integral over all Hilbert spaces $\mathcal{H}^{\nu; 0,0}\equiv \mathcal{H}^{\nu}$ carrying  the principal series scalar representations $U^{(\nu,0)}\equiv U^\nu$, considering that $U^\nu$ and $U^{-\nu}$ are equivalent \cite{tak,moy}:
\b \label{ophss} \mathcal{H}^{(1)}\equiv \int_0^\infty \ud\nu\,\varrho(\nu)\;\mathcal{H}^{\nu} \, .\e
Here $\varrho(\nu)$ is a positive weight for the scalar field in the dS background \cite{brmo96}.
 
To build a (continuous-discrete) Hilbertian basis for a massive scalar field, let us choose $ Q_0^{dS}, \hat{L}^2, \hat{J}^2$, and $\hat{J}_3$ as forming a maximal set of commuting self-adjoint operators representing the dS enveloping algebra and acting in the Hilbert space $L^2(X,\ud\mu(x))\equiv L^2_X$ equipped with the O$(1,4)$ invariant measure on the dS hyperboloid \eqref{dSs}. $Q_0^{dS}$ is the second-order scalar Casimir operator of the dS group SO$_0(1,4)$ and $\hat{L}^2$ is the second-order scalar Casimir operator of the SO($4$) group. Since these operators are self-adjoint, their eigenvalues are real, and their common eigenvectors (in the distributional sense) 
 \begin{equation}
\label{tensnu}
|\nu\,;L,l,m_l \rangle=|\nu\, \rangle \otimes |L,l,m_l \rangle
\end{equation}  
form an orthonormal basis for the Hilbert space of scalar field states. We have \cite{engapewa,moy}:
\begin{align}
\label{l2ket}
    \hat{L}^2|\nu\,;L,l,m_l \rangle &=L\left( L+2\right) |\nu\,;L,l,m_l \rangle\,,\;\; L\in \N\,,   \\
 \label{J2ket}   \hat{J}^2|\nu\,;L,l,m_l \rangle &=l\left( l+1\right) |\nu\,;L,l,m_l \rangle\,,\;\; l=0,1,2, \dots, L\,,\\
 \label{J3ket}   \hat{J}_3|\nu\,;L,l,m_l \rangle &=m_l |\nu\,;L,l,m_l \rangle\,,\;\; -l\leq m_l \leq l\,,  
\end{align}
 \begin{equation}
\label{waveeq1}
Q_0^{dS}|\nu\,;L,l,m_l \rangle = \left( \nu^2+\frac{9}{4}\right) |\nu\,;L,l,m_l \rangle \, ,
\end{equation}
where the parameter $\nu$ assumes the values
\begin{equation}
\label{nuval}
\left\lbrace\begin{array}{cc}
   \nu\in \R^+    &  \mbox{for the principal series,}  \\
- \frac{3}{2}  < \ii\nu <\frac{3}{2}&  \mbox{for the complementary series,}\\
\nu = -\ii\left(n +\frac{1}{2}\right)\, , n\in \N_{\ast} & \mbox{for the discrete series.}
\end{array}\right.\,
\end{equation}

As previously proved in \cite{tak,mol,mol2}, only eigendistributions corresponding to the discrete and principal series are involved in the orthogonal resolution of the identity in $L^2_X$:
\begin{align}
\nonumber
\bu_X&=  \sum_{n,L,l,m_l}c_{n} |-\ii\left(n +\frac{1}{2}\right)\,;L,l,m_l \rangle \langle -\ii\left(n +\frac{1}{2}\right) \,;L,l,m_l| +\\ 
\nonumber &+ \int_0^{+\infty}\ud\nu \,\varrho(\nu)\sum_{L,l,m_l} |\nu\,;L,l,m_l \rangle \langle \nu\,;L,l,m_l |\\
\label{resUL2X} &= \mathbb{P}_{\mathrm{d}} + \mathbb{P}_{\mathrm{c}}\,,
\end{align}
where $\mathbb{P}_{\mathrm{d}}$ (resp. $\mathbb{P}_{\mathrm{c}}$) is the orthogonal projector on the discrete (resp. continuous) set of eigendistributions. The decomposition \eqref{resUL2X} is reminiscent of the decomposition of the identity in $L^2\left(\R^2,\ud^3\mathbf{r}\right)$ into two parts, the discrete part built from the bound states of the H-atom, and the continuous part built from scattering states of the latter \cite{betsal}. Nevertheless, there is a deep difference between both since, in the H-atom case, all bound states are square integrable.  

The orthogonality relations for the discrete \& continuous set of eigendistributions have  the following form:
\b \label{noofba} \langle \nu\,;L,l,m_l |\nu'\,;L',l',m'_l \rangle =
N(\nu,L)\delta(\nu-\nu')\delta_{LL'}\delta_{ll'}\delta_{m_l m'_l}\,,\e
where $N(\nu,L)$ is a normalization factor related to the coefficients $c_{n}$ and weight $\varrho(\nu)$ in \eqref{resUL2X}. Therefore the one-particle Hilbert space  $\mathcal{H}^{(1)}$ introduced in \eqref{ophss} is identified with the closure of $\mathbb{P}_c L^2_X$ and the  identity operator in the Hilbert space  \eqref{ophss}  reads as:
\b \label{idntityco} \bu_{\mathcal{H}^{(1)}}\equiv \mathbb{P}_c =\int_0^\infty \ud \nu\,\varrho(\nu) \sum_{L,l,m_l} |\nu\,;L,l,m_l \rangle \langle \nu\,;L,l,m_l |\,.\e

The eigendistributions which are solutions to \eqref{waveeq1} and involved in the above decomposition are written in dS ambient space formalism by using the so-called dS plane waves as generating functions. They play the r\^ole of Fourier exponentials for the harmonic analysis on the dS hyperboloid.  The latter are also solutions to  \eqref{waveeq1}. They are defined, using the notations of Appendix \ref{notations}, as the
boundary value of the analytic continuation of the functions $\left(H x. \xi \right)^{\sigma}$ to the forward tube $\mathcal{T}^+$ of the complexified dS manifold \cite{brgamo,brmo96,garhuren}:
\begin{align}\label{planket}
 \phi_{\xi\sigma}(x) & \equiv c_{\nu}\mbox{bv}(Hz\cdot\xi)^{\sigma}\\
 &=  c_{\nu} \bigl [\Theta(H x\cdot \xi) + \Theta(-H x\cdot\xi)\,e^{-\ii\pi \sigma }\bigr ] |Hx\cdot\xi|^{\sigma},\; c_{\nu}=\sqrt{\frac{\left(\nu^2+1/4\right)}
{2(2\pi)^3(1+e^{-2\pi\nu})\left(\nu^2+9/4\right)}},\nonumber
\end{align}
where $\xi$ lies in  $C^+$, the upper null cone in the $1+4$ Minkowskian ambient space, $z \equiv x + \ii y$ belongs to  $\mathcal{T}^+$ 
and $\Theta$ is the Heaviside step function. The parameter $\sigma$ is related to the parameter $\nu$ as follows:
\begin{equation}
\label{signu}
\sigma = -\frac{3}{2} - \ii \nu\, , 
\end{equation}
which becomes $\sigma= -n-2$ in the case of discrete series. As generating functions, the dS plane waves admit the following expansion \cite{gasiyo}:
\b \label{depvparv}
\phi_{\xi\sigma}(x)  = 2\pi^2 \left(\xi^0\right)^{\sigma} \sum_{Llm_l} \Phi_{Llm_l}^{\sigma}(x)\mathcal{Y}^{\ast}_{Llm_l}(\pmb{u}) \,,
\e
where $\xi^0>0$, $\xi^\alpha=(\xi^0, \xi^0\; \pmb{u})\in C^+$ and $\pmb{u}=(u^4, \vec u)$ is a unit vector in $\R^4$. Thanks to the Fourier transformation on $\mathbb{S}^3$ based on the orthogonality of the set of hyperspherical harmonics, the expansion \eqref{depvparv} yields the following integral representation \cite{gasiyo},
 \b \label{phix}
 \Phi_{Llm_l}^{\sigma} (x) = \frac{1}{2\pi^2 (\xi^0)^{\sigma} } \int_{\mathbb{S}^3} \ud \pmb{u} \, \phi_{\xi\sigma}(x)  \mathcal{Y}_{Llm_l}(\pmb{u})\,.
 \e
Let us now introduce the so-called $|x\rg$-representation of points in the dS hyperboloid, which makes our construction of the Plancherel formula quite straightforward. First we introduce the Dirac distribution $\delta_P(x)$  on $X$ as having its support at the point $P=(0,0,0,0,H^{-1})$, whose invariance subgroup ($\sim$ stabilizer)  is Lorentz SO$_0(1,3)$:
\begin{equation}
\label{deltaP}
\int_X \ud\mu(x)\, \delta_P(x)=1\quad \mbox{and} \quad \int_X \ud\mu(x)\, \delta_P(x)\, f(x)= f(P)\, , 
\end{equation} 
for all test functions in some dense subspace of $L^2_X$, e.g., infinitely differentiable with compact support.  For any $g\in \mathrm{O}(1,4)$ and from the $\mathrm{O}(1,4)$ invariance of the measure, we have
\begin{equation}
\label{dgP}
 \int_X \ud\mu(x) f(g^{-1}\cdot x)\delta_P(x)= f(g^{-1}\cdot P) = \int_X \ud\mu(x) f(x) \delta_P(g\cdot x)\, . 
\end{equation}
This entails the transformation property of $\delta_P$:
\begin{equation}
\label{covdelt}
 \delta_P(g.x)=  \delta_{g^{-1}\cdot P}(x)\, , 
\end{equation}
which entails the definition of the Dirac distribution $\delta_y$ with support at any point $y\in X$:
\begin{equation}
\label{dely}
\int_X \ud\mu(x) f(x)\delta_y(x) = f(y)\,.  
\end{equation}
Borrowing notations from standard quantum mechanics, we introduce the  set of kets $|x\rg$ and their dual bras $\lg x|$, both labeled by the points  $x\in X$, as obeying the following orthogonality and normalization (in the distributional sense) and resolution of the unity in  $L^2(X)$ 
\begin{align}
\label{normresxx}
\lg x|x^{\prime}\rg&= \delta_{x^{\prime}}(x)=  \delta_{x}(x^{\prime})\equiv \delta(x,x^{\prime})\,, \\ \bu_X&= \int_X\ud\mu(x) \,|x\rg \lg x| 
\end{align}
From its construction, we derive the invariance property of the Dirac distribution on $X$:
\begin{equation}
\label{invdelt}
 \delta(g\cdot x,g\cdot x^{\prime})= \delta(x,x^{\prime})\ \forall g\in \mathrm{O}(1,4)\,. 
\end{equation}
With these notations,  one can now  write
\b \label{ambpar} \Phi_{Llm_l}^{\sigma} (x)\equiv\langle x \vert \nu\,;L,l,m_l \rangle\, .\e

Similarly, we introduce the $|\xi_\gamma\rg$-representation of points in the positive cone $C^+$, or equivalently on the orbital basis $\gamma \sim \mathbb{S}^3$.  The kets $|\xi_\gamma\rg$ and bras $\lg \xi_\gamma|$, both labeled by the points  $\xi\in C^+$, obey the following orthogonality and normalization (in the distributional sense) and resolution of unity in the Hilbert space $L^2(\gamma,\ud\mu_\gamma)\sim L^2(\mathbb{S}^3,\ud\pmb u)$ equipped with the O$(4)$ invariant measure on $\gamma$:
\begin{equation}
\label{identityxi}
\lg \xi_\gamma|\xi_\gamma^{\prime}\rg= \delta_{C^+ }(\xi,\xi^{\prime})\, , \quad \int_\gamma \ud\mu_\gamma(\xi)\, \vert \xi_\gamma \rangle \langle \xi_\gamma\vert= \bu_{C^+ }\, .
\end{equation}
From the equation \eqref{depvparv} and \eqref{ambpar}, in these notations, one can define:
\b \label{xillm} \left< \nu\,; L,l,m_l|\xi_\gamma\right\rangle \equiv 2\pi^2 \big(\xi^0 \big)^{\sigma} \mathcal{Y}^{\ast}_{Llm_l}(\pmb{u})\, ,\e
 Now, by referring to \eqref{tensnu} one formally introduces the kets $|\nu\, ;\xi_\gamma \rangle \equiv |\nu\, \rangle \otimes |\xi_\gamma\rangle $ to give a sense to the expressions
 \b \label{planket0}  \langle x| \nu\,;\xi_\gamma\rangle\,\equiv \phi_{\xi\sigma}(x) \,.\e
Hence we can state the resolution of the identity in the Hilbert space $\mathcal{H}^{(1)}$ defined in \eqref{ophss}, 
\b \label{iderv} \bu_{\mathcal{H}^{(1)}}= \int_0^\infty \ud\nu\,\varrho(\nu) \int_\gamma \ud\mu_\gamma(\xi)\, \vert \nu\;;\xi_\gamma \rangle \langle \nu\,;\xi_\gamma \vert\,. \e 

By replacing the equation \eqref{ambpar} and \eqref{xillm} in the equation \eqref{depvparv}, taking an integral over $\nu$,  and using the identity operator \eqref{idntityco}, one obtains:
\b \label{plaparrr}
\begin{split}
\int_0^\infty d\nu\,\varrho(\nu) 
\phi_{\xi\sigma}(x)
&=\int_0^\infty \ud\nu\,\varrho(\nu)\sum_{Llm_l} \Phi_{Llm_l}^{\sigma} (x)\ 2\pi^2 \big(\xi^0 \big)^{\sigma} \mathcal{Y}^{\ast}_{Llm_l}(\pmb{u})\\
&=\int_0^\infty \ud\nu\,\varrho(\nu)  \sum_{Llm_l} \langle x|\nu\,; L,l,m_l\rangle\langle\nu\,; L,l,m_l|\xi_\gamma \rangle\equiv
 \langle x| \xi_\gamma\rangle\, ,
\end{split}
\e
which determines the relationship between $x$-space and $\xi$-space representations. The above integral \eqref{plaparrr} is well defined and finite (see Lemma $5.1$ in p. $423$ of \cite{tak}).


\subsection{Fock space} \label{B}~

To obtain the Fock space of (scalar) QFT (second quantization) we can start with the following infinite-dimensional closed local algebra:
\b \label{algebcom} [\phi(x), \phi(x')]\equiv G(x,x') \bu \, ,\e
where $G(x,x')=\mathcal{W}(x,x')-\mathcal{W}(x',x)$ is the commutation two-point function, which is zero for space-like separate points. $\mathcal{W}(x,x')=\langle \Omega |\phi(x)\phi(x')|\Omega\rangle$ is the Wightman two-point function and $|\Omega\rangle$ is the vacuum state. By contrast to the first quantization procedure, one cannot 
directly construct the Hilbert or Fock space from this local algebra of free field operators since the limit $x \longrightarrow x'$ is singular as the state is not normalizable in this limit. Therefore the operators must be defined in a tempered-distributional sense on an open subset $\mathcal{O}$ of spacetime \cite{brmo96}:
\b \label{algebcom2} [\phi(f), \phi(g)]\equiv G(f,g) \bu \, ,\e
where $f$ and $g$ are test functions and $G(f,g)=\int \ud\mu(x)\ud\mu(x')f(x)g(x')G(x,x')$.

As usual, we assume that the field operator can be written in terms of its creation part, $\phi^+$, and its annihilation parts $\phi^-$:
$$ \phi(f)=\phi^-(f)+\phi^+(f)\, , $$
where $\phi^+(f)$ creates a state and $\phi^-(f)$ annihilates a state in the considered Fock space. We define a ``number'' operator $N(f,g)$ in the Fock space as follows:
\b \label{nuop} N(f,g)\equiv \phi^+(f)\phi^-(g)\,, \;\;\; \forall \;|\psi\rangle\in \mathcal{H} , \;\; N(f,g)|\psi\rangle \in \mathcal{H}\,.\e
Using the equations $(2.10)$ and $(2.11)$ of \cite{brmo96}, one can calculate the effect of $N(f,g)$ on the Hilbert space.
Similarly, one can prove the following algebra, which results in the construction of the Hilbert space:
\b \label{gcsll} \left\{\begin{array}{clcr} \left[\phi^-(f), \phi^+ (g)\right] &= \mathcal{W}(f,g) \bu \,,\\
\left[N(f,g), \phi^+ (k)\right]&=\mathcal{W}(g,k) \phi^+(f) \, ,\\
\left[N(f,g), \phi^-(k)\right]&= -\mathcal{W}(k,f) \phi^-(g) \, ,
\end{array} \right.
\e
where $\mathcal{W}(f,g)=\int \ud\mu(x)\ud\mu(x')f(x)g(x')\mathcal{W}(x,x')$.

Now using the infinite-dimensional closed local algebra \eqref{gcsll}, one can construct the Hilbert space in a distributional sense on an open subset $\mathcal{O}$ of the dS spacetime \cite{brmo96}. The action of the annihilation part of the field operator in the vacuum or ground state, $|\Omega\rangle $, yields zero, and the creation part produces the one-particle state:
$$\phi^-|\Omega\rangle= 0,\;\;\; \phi^+|\Omega\rangle= |\psi_1\rangle \in \mathcal{H}^{(1)}\,.$$
This construction consists of two main steps. The first amounts to fixing the vacuum state's norm, which has been done in the null curvature limit \cite{brmo96}. The second is to identify the one-particle Hilbert space with that used in the first quantization \eqref{ophss}. Then the Fock space can be 
defined as usual as the Hilbertian sum:
\begin{equation}\label{DEF-FockSpace} \mathcal{F}\equiv \mathcal{H} \equiv \left\{ \C, \mathcal{H}^{(1)}, \mathcal{H}^{(2)},\cdots, \mathcal{H}^{(\mathsf{n})}, \cdots \right\}, \end{equation}
where $\C$ is spanned by the vacuum state, $\mathcal{H}^{(1)}$ is the space of one-particle states, and $\mathcal{H}^{(\mathsf{n})}$ is the space of $\mathsf{n}$-particles states. 
This $ \mathsf {n}$-particle states are built
through symmetric tensor products of one-particle states in our scalar, thus bosonic, case. The field operator 
appears as a map on Fock space $ \phi : \mathcal{F} \longrightarrow \mathcal{F}\,$.

Finally, let us consider the in and out Hilbert or Fock spaces: $\mathcal{F}_{in}$ (resp. $\mathcal{F}_{out}$). 
They consist of free-field states in the gravitational background, which do not interact with themselves or other quantum fields. As we shall discuss in the following, due to the homogeneity of the dS spacetime they can be defined through all dS spacetime from the free field states with an integral over $\nu$.


\section{Asymptotic completeness theorem} \label{asycop}

To address the question of the existence of asymptotic states and their completeness, let us first briefly review and comment on three central results, namely: the Plancherel theorem, the Reeh-Schlieder theorem, and the Kall\'en-Lehmann type representation in Minkowski (see App. \ref{ap}) and dS spaces. Then, we will use them to establish the asymptotic state completeness theorem in dS space.

\subsection{Existing results}
The Plancherel theorem states that the integral of the squared modulus of a function in Minkowski spacetime is equal to the integral of the squared modulus of its energy-momentum Fourier counterpart (see App. \ref{ap}). In \cite{tak}, Takahashi established the Plancherel formulas for the dS group SO($1,4$); namely, the integral of the squared modulus of a function on the group parameters with its Haar measure is equal to the integral of the squared modulus of its ``spherical transform'' on the part of the unitary dual of the group equipped with a specific measure. Next, Molchanov used the Takahashi results to define the Plancherel formulas between two different homogeneous spaces \cite{mol2}. In the $2$-dimensional dS spacetime, Bros et al. constructed the Plancherel formula \cite{brmo2} between a space of functions on the dS spacetime (actually its complexified version $\mathcal{T}^\pm)$ \eqref{tpm} and space of functions defined on $ R^+_{(\nu)} \times C^+$ (real positive line $+$ positive null cone). For any pair of functions $(f,g)$ in the $2$-dimensional dS version of $\mathcal{H}^{(1)}$ and their corresponding decompositions, one has (see Appendix \ref{apc}) \cite{brmo2}:
\b \label{plan} \int_X \ud\mu(x) \overline{f(x)} g(x) =\frac{1}{2\pi^2} \int_0^\infty \ud\nu \; \frac{\nu \tanh \pi\nu }{e^{\pi\nu}\cosh \pi\nu}\int_\gamma \ud\mu_\gamma(\xi) \overline{\tilde{f}(\xi,\nu)}\tilde{g}(\xi,\nu) \,.\e
As explained in the introduction, the parameter $\nu \in \R^{+}$ labels the representation of the dS group in the principal series with spin $0$ and is viewed as a dS ``rest energy'' \eqref{mEdS}.

The Bros-Fourier-Helgason transform, in $4$-dimensional dS spacetime, can be defined in terms of the dS plane waves. Here for the massive scalar case, by using the equations \eqref{planket} and \eqref{normresxx}, we obtain \cite{brmo96}:
\b \label{dsfou} \mathcal{H}^{(1)}\ni f(x) \mapsto \langle \nu\,;\xi_\gamma| f\rangle\equiv \tilde f(\xi,\nu)=\int_X \ud\mu(x)\,\left\{\phi_{{\xi \sigma=-\frac{3}{2}+\ii\nu}}(x) f(x) \right\}\, .\e 
For obtaining the inverse Bros-Fourier-Helgason transform, we must define the Dirac delta function in the ambient space formalism. Within this context, an analogous of the Dirac delta distribution can be introduced from Cauchy's integral type representation, see equations \eqref{delt2} and \eqref{delt4}. Therefore, the inverse Bros-Fourier-Helgason transform can be written in the following form:
\b \label{indsfou} \langle x|f\rangle\equiv f(x)=\int_0^\infty \ud \nu\, \varrho(\nu) \; \int_\gamma \ud\mu_\gamma(\xi)\ \left\{\phi_{{\xi \sigma=-\frac{3}{2}-\ii\nu}} (x)\tilde f(\xi,\nu)\right\},\e
see Appendix \ref{apc}, and more details are given in \cite{brmo2}. By replacing the identity operators \eqref{normresxx} and \eqref{iderv} in $\langle f|g\rangle$, the Plancherel formula in $4$-dimensional dS spacetime can be obtained for all $f,g \in \mathcal{H}^{(1)}$:
\b \label{plan4d} \lg f|g\rg_{L^2_X} =\int_X \ud\mu(x) \overline{f(x)} g(x) =\int_0^\infty \ud\nu \; \varrho (\nu)\int_\gamma \ud\mu_\gamma(\xi) \overline{\tilde{f}(\xi,\nu)}\tilde{g}(\xi,\nu) \,.\e
For the mcc scalar field in $4$-dimension see Appendix \ref{apd}.

The Reeh-Schlieder theorem states that the vacuum state $\vert \Omega \rangle$ is a cyclic vector for the field algebra $\mathcal {A}(\mathcal{O})$, $\mathcal{O}$ is some open set of the Minkowski spacetime.
Any state $\vert \psi \rangle$ can be approximated to arbitrary precision by acting on the vacuum with an operator selected from local algebra. This is even for states $\vert \psi \rangle$ that contain excitations arbitrarily far away in space. In this sense, the quantum states created by applying elements of the local algebra to the vacuum state are not localized to the region $\mathcal{O}$ (see App. \ref{ap}) \cite{wit}. The Reeh-Schlieder property for generalized free fields in dS spacetime was proved by Bros \textit{et al} \cite{brmo96,brepmo}.

The Kall\'en-Lehmann spectral representation in Minkowski spacetime gives a general expression for the (time-ordered) two-point function of an interacting quantum field theory as a sum of free propagators with different mass parameters (see App. \ref{ap}). Bros et al. proved the Kall\'en-Lehmann type representation in dS spacetime \cite{brmo96}. By replacing the spectral condition with the geodesic spectral condition, they defined the two-point function through
\b \label{kale} \mathcal{W}(x,x')=\int_0^\infty \ud \nu\,\varrho(\nu)\; \mathcal{W}_\nu(x,x')\, , \e
where $\varrho(\nu)$ is now a positive measure for the interacting field theory in the dS background, which is different from the dS free field theory, see \cite{brepmo2}.

\subsection{Proof} 
Under an extension of the adiabatic hypothesis (see Eq. \eqref{adia} below),  we present proof of the asymptotic state completeness theorem. This proof amounts to show the following: 
\begin{enumerate}
 \item one-particle states completeness,
 \item existence and completeness of asymptotic states,
 \item generalization to the theory of the interacting field.
\end{enumerate}
\par
The first part comes out from the Plancherel formula in dS space. The equation \eqref{plan4d} is globally defined on the dS hyperboloid and is independent of a given observer. A choice of the latter would be linked to a cosmological event horizon, and a global definition would be lost. Contrary to the situation in Minkowski space,  one cannot define the inverse of the Bros-Fourier-Helgason transform in dS spacetime globally by using the quantum states for a single value of the parameter $\nu$. In effect, these
one-particle free states with a definite parameter $\nu$ cannot form a complete orthogonal family of states or basis in the space of states on dS space: due to the interaction with the background geometry, 
we have particle creation and entanglement of different number states. In this case, a complete ``one-particle state'' should be defined as an integral over the parameter $\nu$; see Section \ref{A}. The Hilbert space 
$\mathcal{H}^{\nu}$ may be realized as the space $L^2 (\gamma,\ud\mu_\gamma) \sim L^2(\mathbb{S}^3,\ud\pmb u)$ of complex-valued functions defined on the orbital basis $\gamma \sim \mathbb{S}^3 \subset \C^+ $, then the one-particle Hilbert space of the theory $\mathcal{H}^{(1)}$ can be defined by:
\b \label{ther1}  \mathcal{H}^{(1)}=\int_0^\infty \ud \nu\, \varrho_0(\nu) \; \mathcal{H}^{\nu}\, , \e
where $\varrho_0(\nu)\,\ud \nu$ is a positive measure for the free field in a dS background gravitational field, {\it i.e.} the interaction with the dS background gravitational field is only considered.

It is important to note that the function $f(x)$ 
in Eq. \eqref{indsfou} is not a solution of the homogeneous dS-Klein-Gordon equation \eqref{waveeq1} with a specific parameter $\nu$. The non-homogeneous part of the dS-Klein-Gordon 
equation of  $f(x)$ is due to the interaction with the classical gravitational dS background viewed as a source. It is apparent from Eq. \eqref{indsfou} that since we take the integral over $\nu$, the concept of a particle associated with a specific value of the parameter $\nu$ is meaningless for dS QFT. We have a superposition of different $\nu$-states due to the interaction between the classical gravitational background and the field operators. In curved spacetime, the effect of gravity cannot be turned off. It appears as an unavoidable classical background source.

The second part is concerned with asymptotic states. They exist since the field operators 
are defined globally on the dS hyperboloid: from the algebra \eqref{gcsll} and the definition \eqref{DEF-FockSpace}, one can see that the vacuum state is cyclic under the action of a local field algebra $\mathcal{A}$, {\it i.e.} all states can be constructed by applying the algebra of field operators to the vacuum state. That means for each local operator $A(f)\in \mathcal{A}$, we have
\b \label{the2} A(f_1,\cdots, f_n) |\Omega\rangle =|\psi(f_1,\cdots, f_n)\rangle\in \mathcal{F}\,.\e
Since the de~Sitter spacetime is a homogeneous space, this construction extends to
each open set around each spacetime point. This argument is precisely formalized by the Reeh-Schlieder property extended to dS spacetime. It applies, in particular to the future and the past infinity limits (for definition see next section), which, in turn, ensure the existence of the asymptotic states, $in$ and $out$. These states form a complete family of states corresponding to the asymptotic (Fock) spaces: $\mathcal{F}_{in}$, and $\mathcal{F}_{out}$. Moreover, this construction implies that $\mathcal{F}_{in}\equiv \mathcal{F}_{out}$. 

It is instructive to note that in Minkowski's space, local algebra can also be generalized to all of the points of spacetime by using the Bisognano-Wichmann theorem, which describes the relationship of local algebra with the Poincar\'e group \cite{biwi}. The Bisognano-Wichmann theorem has also been proved in the dS spacetime \cite{brepmo}. Then the dS Symmetry requires that the action of SO$(1,4)$ group maps the local algebra of one region onto the algebra of the transformed region, thereby preserving all algebraic relations and then the Hilbert space structure.

The third part of the proof is concerned with the interacting field. In the general interacting case, one can define the one-particle Hilbert space similarly to the direct integral on the Hilbert space carrying the massive scalar dS UIRs:
\b \label{onpahles1} \mathcal{H}^{(1)}_I=\int_0^\infty \; \ud \nu\, \varrho(\nu) \; \mathcal{H}^{\nu}\equiv \int_0^\infty \; \ud \nu\,\varrho(\nu)\int_\gamma \ud\mu_\gamma(\xi) \; \mathcal{H}_{\xi}^{\nu}\, , \e
where $\varrho(\nu)$ is a positive measure for the interacting field theory. Since the Hilbert space, $\mathcal{H}_{in}^{(1)}$ and $ \mathcal{H}_{out}^{(1)}$ are complete, {\it i.e.} there exists the identity operator \eqref{iderv}, then we have $\mathcal{H}_{in}^{(1)}\equiv \mathcal{H}^{(1)}_I\equiv \mathcal{H}_{out}^{(1)}$. Now it must be generalized to the interaction QFT.

Let us assume all interactions vanish at the limit of future and past infinity ($x^0 \longrightarrow \pm \infty$), except for the classical background gravitational interaction. The physical meaning of the limit of future and past infinity will be presented in the next section. Likewise, in the Minkowski spacetime \cite{itzu}, we use the adiabatic hypothesis conditions in the interaction case within the framework of dS ambient space formalism. All physical quantities must be understood at the limit when this adiabatic switching is removed. We extend the adiabatic hypothesis in dS space through the following form:
\b \label{adia} x^0 \longrightarrow -\infty\,, \;\;\;\; \phi(x) \longrightarrow \int_0^\infty \ud\nu\,\varrho_a(\nu)\,  Z^{\frac{1}{2}} \phi_{in}^\nu(x)\,,\e
where $Z$ is a c-number (not an operator), and $ \phi_{in}^\nu(x) $ is a free field operator. For the various examples
of $ \varrho_a(\nu)$ in the adiabatic limit in dS space-time, see Section $5$, appendices B and C in \cite{brepmo2}. Here the explicit form of it does not change our conclusion, and we use only its positivity properties. In an ideal simple case, it can be thought of as a form of the delta function $\delta(\nu-\nu')$, but in a full dS situation, we have a spectrum of $\nu$ due to the creation of particles from the gravitational background field \cite{brepmo2}.

Using the equation \eqref{kale}, the commutation two-point function read as:
\b \label{the4} G(x,x')=\langle \Omega |[\phi(x),\phi(x')]|\Omega\rangle= \int_0^\infty \; \ud \nu\,\varrho(\nu)\;  G(x,x';\nu)\, . \e
Using the equation \eqref{adia} and \eqref{the4}, and from the renormalizability of $\phi_{in}^\nu(x)$ in $G(x,x';\nu)$, the positiveness of $\varrho(\nu)$ (Theorem $5.1$ Ref. \cite{brmo96}) and convergence property of the integral \eqref {the4} (Lemma $5.1$ in page $423$ of \cite{tak} 
and Theorem $2$ Ref. \cite{brmo2}), we can conclude that $0\leq Z<\infty$. Through its renormalization, one can write $0\leq Z<1$, then $\phi(x)$ is also renormalizable. Therefore one can identify the free and interaction Hilbert (Fock) spaces since they are complete:
\b \label{asscomp} \mathcal{F}_{in}\equiv \mathcal{F}\equiv \mathcal{F}_{out}\, .
\e


\section{Future and past infinity limit} \label{timevop}

With ambient space coordinates $(x^\alpha)\equiv (x^0, \pmb x)$, $x^0$ plays the same role as time as in Minkowski space. In global coordinate systems, $x^0$ is in one-to-one correspondence with the coordinate $\vartheta=Ht$ or the conformal time $\rho=H\tau$ (see Eq. \eqref{gcs}), and $t$ or $\tau$ can be viewed as the Minkowskian time through a group contraction process at the null curvature limit. Note that the sign of $t$ or $\tau$ cannot change the sign of $\,\pmb x$. The variables $t$, $\tau$, and $x^0$ increase together but at different rates. The coordinate $x^0$ can provide a time direction in this coordinate system on the dS hyperboloid. Therefore, in the ambient space formalism, choosing $x^0=\infty$ ($x^0=-\infty$) as the future (past) limit of the quantum state makes sense, although the time translation symmetry property does not exist.

In the previous section, we proved the completeness theorem in dS QFT using an extension of the adiabatic hypothesis. This hypothesis is based on the definition of a quantum state in the future and past infinity limit: $x^0 \longrightarrow \pm \infty$. When the dS group generators $J_a$ and field operators $\phi$ 
are applied to quantum states in Hilbert space, a new state in Hilbert space is created:
\b \label{evolution} |\psi\rangle \in \mathcal{H} \Longrightarrow U(J_a\,,\phi)|\psi \rangle=|\psi' \rangle \in \mathcal{H}\, ,\e
 where $U$ is a well-defined function of the dS group generators and the field operators. In this section, we would like to consider the relationship between two limited quantum states in the ambient space formalism {\it i.e.} $|\Psi\,;x^0=\infty\rangle$ and $|\Psi\, ; x^0=-\infty\rangle$. Then, there relation to
the existence of the $S$-matrix operator from an observer-independent viewpoint is discussed.
In the ambient space formalism, the retarded propagator is defined as \cite{brmo96}:
\b \mathcal{R}(x,x')=\theta(x^0-x'^0)G(x,x')\,, \e
where $G(x,x')$ is the commutation function. $G(x,x')$ is dS invariant and independent of the choice of the vacuum state. It is a well-defined function at the infinity limit of the future and past. Therefore, we can define two sets of quantum states in the limit of future and past infinity,
that belong to their corresponding Hilbert spaces:
\b \label{teod} \vert \psi\,,x^0=-\infty\rangle\equiv \vert \psi\rangle_{in} \in \mathcal{H}_{in}\, \;,\;\;\; \vert \chi\,,x^0=\infty\rangle\equiv \vert \chi\rangle_{out}\in \mathcal{H}_{out} \, .\e
Since these two spaces are equivalent and complete, there exists a unitary operator, which maps states in one to states in the other one:
$$ \mathcal{U}(in\,,out) \vert \psi\rangle_{in} =\vert \chi\rangle_{out}\,.$$

The operator $\mathcal{U}(in\,,out)$ should be a well-defined function of the dS group generators $M_{\alpha\beta}$ and the field operators $\phi$ similar to the equation \eqref{evolution}. The dS group generators $M_{\alpha\beta}$ act on the vector state in Hilbert space, which can be defined in all of the points of spacetime, and change it to the new state in the one-particle Hilbert space. The creation and annihilation of particles are defined by the field operators $\phi$. The field operator $\phi$, on the one hand, is a map on the Fock space, and on the other hand, it is defined in spacetime geometry. The deep connection between the algebraic structure (Fock or Hilbert space $\mathcal{H}$) and the geometry of spacetime (manifold $X$) can be extracted from the Bisognano-Wichmann theorem. For the scalar Wightman field, $\phi$ in Minkowski space see \cite{mori}.

Since in the null curvature limit $\vert \psi\rangle_{in}$ and $\vert \chi\rangle_{out}$ became the states in Minkowski space, the operator $\mathcal{U}(in\,,out)$ can be identified with the $S$-matrix operator in Minkowski spacetime in this limit. Therefore we define the ``$S$-matrix operator" in dS ambient space formalism by this operator: $S\equiv\mathcal{U}(in\,,out)$. Can we claim that this definition is dS invariant?

In dS classical field theory, the Lagrangian density of free fields can be obtained from the free field equation, which can be calculated from the second-order Casimir operator of the dS group \cite{ta1}. The interacting fields can be
formulated through a gauge transformation. The total Lagrangian density can be calculated in the dS ambient space formalism with the gauge fixing, and the ghost fields \cite{taga}. In this case, the Lagrangian density is a function of the parameter $\nu$, $\mathcal{L}(\Phi,\partial^\top_\alpha \Phi,\nu)$. On the quantum level, one cannot calculate the dS invariant $S$-matrix elements from this Lagrangian density since, due to the classical background gravitational field, we have the creation of particles with different values of $\nu$. Furthermore, an integral over the parameter $\nu$ is needed to prove the asymptotic state completeness.

Let us go back to the property of completeness in QFT exemplified by the equation \eqref{onpahles1}. In classical field theory, a scalar field with a specific parameter $\nu$ can be defined as the solution to the dS-Klein-Gordon field equation $$\left(Q_0^{dS}-\left(\nu^2 + \frac{9}{4}\right)\right)\phi_\nu(x)=0\, .$$
Nevertheless, in QFT, due to the classical background interaction with the field operator and then particle creation, we have a spectrum $\Delta \nu$, whose range depends on the curvature of spacetime. Therefore the quantum field operator is a function of this spectrum, and it can be formally written in the following form in the first approximation:
\b \phi(x) \approx \int_0^\infty \ud \nu \, \varrho(\nu) \, \phi_\nu(x)\, , \;\;\mbox{where}\;\; Q_0^{dS}\phi(x)=j(x)\,.\e
This quantum field operator is not the solution to the homogenous dS-Klein-Gordon field equation. The source $j(x)=\int_0^\infty \ud \nu \, \varrho(\nu) \, \lg Q_0^{dS}\rg\phi_\nu(x)$ determines the interaction between the scalar field $\phi_\nu$ with the dS geometry, $\lg Q_0^{dS}\rg$, {\it i.e.} the background gravitational field.
Then the Lagrangian density must be replaced in the following way:
\b \mathcal{L}_c(\Phi,\partial^\top_\alpha \Phi,\nu)\;\; \longrightarrow \int_0^\infty \ud \nu \, \varrho(\nu)\; \mathcal{L}_c(\Phi,\partial^\top_\alpha \Phi,\nu) \, .\e

With this replacement, the definition of dS invariant S-Matrix elements is principally possible in ambient space formalism, {\it i.e.} for an observer-independent point of view. This replacement for an observer may be interpreted as the creation of particles due to the universe's expansion or a thermal effect of the vacuum state. It is crucial to assert once more that the integral is due to the impact of the classical background gravitational field on the scalar field operator. The source of particle creation comes from the curvature of spacetime, and the curvature is reduced. Due to the back-reaction of the created particles, the curvature of spacetime is reproduced. It gives rise to energy exchange between geometry and quantum fields, and because of this energy fluctuation, we do not have a global definition of the energy concept in dS space. Note that this effect that we have explained here can be generalized to other spin fields.

\section{Conclusion} \label{conclu}

A dS QFT can be constructed similarly to its Minkowski counterpart on a rigorous mathematical basis by taking advantage of the symmetry properties of dS spacetime and ambient space formalism. Two orthonormal bases (in the distributional sense) for the Hilbert space of de Sitter group algebra are presented for a scalar field. Using the Plancherel formulae, the Reeh-Schlieder theorem,
the Kall\'en-Lehmann type representation, and an extension of the adiabatic limit, we have shown that the asymptotic Fock space exists and it is complete. The Plancherel formulae are used to prove the completeness of the one-particle states, and the Reeh-Schlieder property results in the existence and completeness of asymptotic free states and the Kall\'en-Lehmann type representation proves that the interaction fields do not change the situation. It is important to note that a mass-like spectrum is needed for a free QFT in dS space, similar to the interacting field theory in Minkowski space.

The quantum states of the future and past infinity limits have been defined. Finally, the existence of a dS invariant $S$-matrix element is discussed in ambient space formalism, {\it i.e.} from an observer-independent point of view. The dS completeness theorem of Hilbert space structure allows us to study quantum dS geometry from the fundamental field operator algebra perspective \cite{ta223}, which will be considered in future work \cite{tafinal}.

\vspace{0.5cm}

{\bf{Acknowledgements}}: The author wishes to express his special thanks to Jean Iliopoulos for suggesting this problem and for his discussion. One of the authors, MVT, would like to thank Coll\`ege de France, Universit\'e Paris Cit\'e, and the laboratory APC for their hospitality and financial support.

\begin{appendix}


\section{Notations} \label{notations}

The dS spacetime can be identified with the $4$-dimensional hyperboloid embedded in the $5$-dimensional Minkowski spacetime as:
\b \label{dSs} X=\left\{ x^\alpha \in \R^5| \; \; x \cdot x=\eta_{\alpha\beta} x^\alpha
x^\beta =-H^{-2}\right\}\,,\;\; \alpha,\beta=0,1,2,3,4\,, \e
with $\eta_{\alpha\beta}=$diag$(1,-1,-1,-1,-1)$. The dS metric element reads:
\b \label{dsmet} \ud s^2=\left.\eta_{\alpha\beta}\ud x^{\alpha}\ud x^{\beta}\right|_{x\cdot x=-H^{-2}}=
g_{\mu\nu}^{dS}\ud X^{\mu}\ud X^{\nu}\,,\;\; \mu=0,1,2,3\,,\e
where the $X^\mu$'s form a set of $4$-space-time intrinsic coordinates on the dS hyperboloid, and the $x^{\alpha}$'s are the ambient space coordinates. The ambient space coordinates are conveniently expressed in terms of the global intrinsic coordinates  $(X_s^\mu)=(\vartheta=Ht,\chi,\theta,\varphi)$, $\vartheta\in\R$, $0\leq\chi,\theta\leq\pi$, $0\leq\varphi<2\pi$, or the so-called conformal ones, 
$(X_c^\mu)=(\rho=H\tau,\chi,\theta,\varphi)$, $-\frac{\pi}{2}<\rho<\frac{\pi}{2}\in\R$, as:
\b \label{gcs} \left\{\begin{array}{clcr} x^0&=H^{-1}\sinh \vartheta = H^{-1}\tan\rho\,, \\
x^1&=H^{-1}\cosh \vartheta\sin\chi \cos\theta = H^{-1}\sec\rho \sin\chi \cos\theta\,, \\
x^2&=H^{-1}\cosh \vartheta\sin\chi \sin\theta\cos\phi = H^{-1}\sec\rho\sin\chi \sin\theta\cos\phi\,, \\
x^3&=H^{-1}\cosh \vartheta\sin\chi\sin\theta\sin\phi = H^{-1}\sec\rho\sin\chi\sin\theta\sin\phi\,,  \\
x^4&=H^{-1}\cosh \vartheta\cos\chi= H^{-1}\sec\rho \cos\chi \, .
\end{array} \right.\e
These coordinate systems are suitable for our discussion. For $t$ (resp. $\rho$ or $\tau$) constant, the spatial part is the $3$-sphere with radius $H^{-1}\cosh \vartheta$ (resp. $H^{-1}\sec\rho$).

The analyticity properties of QFT in dS space are understood through the complexified de Sitter spacetime $X^{(c)}$:
$$ X^{(c)}=\left\{ z=x+\ii y\in \C^5;\;\;\eta_{\alpha \beta}z^\alpha z^\beta=(z^0)^2-\vec z.\vec z-(z^4)^2=-H^{-2}\right\}$$
\b =\left\{ (x,y)\in \R^5\times \R^5;\;\; x^2-y^2=-H^{-2},\; x.y=0\right\}\,.\e
The forward and backward tubes in $ \C^5$ are defined as $T^\pm= \R^5+\ii V^\pm$. The domain $V^+$(resp. $V^-)$
stems from the causal structure on $X$:
\b V^\pm=\left\{ x\in \R^5;\;\; x\cdot x=-H^{-2},\;\;\;x^0 \gtrless \sqrt {\parallel \vec x\parallel^2+(x^4)^2}\right \}\,.\e
The forward and backward tubes of the complexified dS space are defined by:
\b \label{tpm} \mathcal{T}^\pm=T^\pm\cap X^{(c)}\,.\e
Finally the ``tuboid'' above $X^{(c)}\times X^{(c)}$ is defined by \cite{brmo96}:
\b \label{tuboid} \mathcal{T}_{12}=\{ (z,z');\;\; z\in \mathcal{T}^+,z' \in \mathcal{T}^- \}\,. \e
The positive light cone is defined as: $ C^+ = \{ \xi \in \R^5;\;\; \xi\cdot \xi=0,\;\; \xi^0 \geq 0, \;\; \xi^\alpha \neq 0 \}$.

An important tool of ambient space formalism is the tangential derivative:
\begin{equation}
\label{tangder}
\partial_\beta^\top =\theta_{\alpha \beta}\partial^{\alpha}=
\partial_\beta + H^2 x_\beta x\cdot\partial\,,
\end{equation}
where $ \theta_{\alpha \beta}=\eta_{\alpha \beta}+
H^2x_{\alpha}x_{\beta}$ is the transverse projector. As a differential operator acting on scalar fields, the second-order Casimir operator $Q_0^{dS}$ of the dS group SO$_0(1,4)$ is written in terms of the tangential derivative as \cite{ta1}: 
\begin{equation}
\label{casQ0}
 Q_0^{dS}=-\frac{1}{2} M_{\alpha\beta}M^{\alpha\beta}=-H^{-2}\partial^\top\cdot\partial^\top=-H^{-2}g^{dS}_{\mu\nu}\nabla^\mu\nabla^\nu=-H^{-2} \large{\Box}_H \,,
\end{equation}
where $ {\Box}_H$ is the Laplace-Beltrami operator on dS space-time and $M_{\alpha\beta}=-i\hbar \left(x_\alpha \partial^\top_\beta-x_\beta\partial^\top_\alpha\right)$.

The condition of transversality $ x\cdot K(x)=ًًًًً0 $ restricts the five-vector fields to the dS hyperboloid and guarantees that they are viewed as vector-valued homogeneous functions of the $\R^5$-variables $x^{\alpha}$ with some arbitrarily chosen degree $\sigma$ \cite{dir}:
\b \label{hode} x^{\alpha}\frac{\partial }{\partial
x^{\alpha}}K_{\beta}(x)=x\cdot \partial\, K_\beta (x)=\sigma
K_{\beta}(x)\,. \e


\section{An orthonormal basis for Klein-Gordon inner product} \label{L2X}

Let us equip the vector space of solutions to the eigenvalue equation \eqref{waveeq1}  with the following Klein-Gordon inner product:
\begin{equation}
\label{kgie}
 \langle\Phi_1,\Phi_2\rangle_{\mathrm{KG}}\equiv
\ii\int_\Sigma \ud\sigma^\mu \,\Phi_1^*\stackrel{\leftrightarrow}{
   \partial}_{\mu}\Phi_2\,,
\end{equation}
where $\Sigma$ is a Cauchy surface, {\it i.e.} a space-like surface such that the Cauchy
data on $\Sigma$ define uniquely a solution of (\ref{waveeq1}), and $\ud \sigma^\mu$
is the area element vector on $\Sigma$. Labeling a point $x \in X$ on dS hyperboloid by the above conformal  coordinates in \eqref{gcs}, namely
\b \label{coglco} (x^\alpha) \equiv (x^0, \pmb x)=(H^{-1}\tan \rho, (H\cos\rho)^{-1}\pmb{ v})\, ,\e
where $\pmb{x} =( \vec x, x^4)$ and  $\pmb{ v} =( \vec v, v^4)$ is a unit vector in $\R^4$, \textit{i.e.} $\pmb{ v} \in \mathbb{S}^3$, the Klein-Gordon  product (\ref{kgie}) can be written
 as
\renewcommand{\arraystretch}{0.6}
\begin{equation}\langle\Phi_1,\Phi_2\rangle_{\mathrm{KG}}=\frac{\ii}{H^2}\int_{\mathbb{S}^3,\rho=0}\! \ud \pmb{ v}\,
 \Phi_1^{*}(\rho,\pmb{ v})\stackrel{\leftrightarrow}{
\partial}_{\rho}\Phi_2(\rho,\pmb{ v}) . 
\end{equation} 
where $\ud \pmb{ v}=\sin^2\alpha\,\sin\theta\,\ud\alpha\,\ud\theta\,\ud\phi$ is
the O($4$) invariant measure on $\mathbb{S}^3$. This product is dS invariant and independent of the choice of $\,\Sigma$. 
Due to the orthogonality of the hyperspherical harmonics, the orthogonal  set of functions $\Phi_{Llm}^{\sigma} (x) $ corresponding to the kets $|\nu\,;L,l,m_l \rangle$ read as \cite{gasiyo}:
\b \label{maelxh}
\begin{split}
 \Phi_{Llm_l}^{\sigma} (x) & = \ii^{L-\sigma}\, e^{-\ii(L+\sigma +3)\rho}(2 \cos{\rho})^{\sigma + 3} \frac{\Gamma(L-\sigma)}{(L+1)!\Gamma(-\sigma)}
\\ & \times {}_2F_1\big(\sigma + 2, L+\sigma +3; L+2;-e^{- 2\ii\rho}\big) \mathcal{Y}_{Llm_l}(\pmb{v})\, ,
\end{split}
\e
where ${}_2F_1$ is the hypergeometric function and $\sigma$ is a complex parameter assuming the value $\sigma= -\frac{3}{2}- \ii\nu$ (resp. $\sigma= -\frac{3}{2}- \nu$, $-3/2 <\nu<3/2$) (resp. $\sigma= -p-2$, $p=1,2,\cdots$) for the scalar principal (resp. complementary) (resp. discrete) series of the dS group SO$_0(1,4)$. The hyperspherical harmonics $ \mathcal{Y}_{Llm_l}$ on the $3$-sphere $\mathbb{S}^3$ are in one-to-one correspondence with the kets $ |L,l,m_l \rangle $. In this sense, they form an orthonormal basis for the Hilbert space $\mathcal{K}\sim L^2(\mathbb{S}^3)$.

Orthogonality relations and norms are given by \cite{gasiyo}:
\begin{equation*}
\langle\Phi_{Llm_l}^{\sigma},\Phi_{L'l'm'_l}^{\sigma}\rangle_{\mathrm{KG}} = \delta_{LL'}\, \delta_{l'l'}\, \delta_{m_lm'_l}\, \Vert \Phi_{Llm_l}^{\sigma}\Vert^2_{\mathrm{KG}}\, , 
\end{equation*}
 where
\begin{align} \label{normPhicompl}
\Vert \Phi_{Llm_l}^{\sigma}\Vert^2_{\mathrm{KG}}&= \frac{ \pi\, 2^{2-2L}}{H^2}e^{-\pi \mathrm{Im}\sigma}\left\vert \frac{(\Gamma(L-\sigma)}{(\Gamma(-\sigma)}\right\vert^2 \times\\
 &\mathrm{Re}\left\lbrack \Gamma^{\ast}\left(\frac{L-\sigma+1}{2}\right)\Gamma\left(\frac{L-\sigma}{2}\right)\Gamma^{\ast}\left(\frac{L+\sigma+4}{2}\right)\Gamma\left(\frac{L+\sigma+3}{2}\right)\right\rbrack^{-1}\,.\nonumber
\end{align}
$\mathrm{Re}$ is the real part and $\mathrm{Im}$ is the imaginary part.


\section{Theorems in the Minkowskian case} \label{ap}

The Plancherel formulae, Reeh-Schlieder theorem, and Kall\'en-Lehmann type representation in Minkowskian QFT are briefly recalled here. The Plancherel theorem states that the integral of the modulus squared of a function is equal to the integral of the modulus squared of its frequency spectrum:
$$ \int _{-\infty }^{\infty } \ud x\,|f(x)|^{2}=\int _{-\infty }^{\infty } \ud k \,|\widehat {f}(k )|^{2}\, .$$
where $f(x)$ is a function on the real line, and $\widehat{f}(k )$ is its frequency spectrum.
Plancherel's theorem remains valid when stated in n-dimensional Euclidean space $\mathbb{R} ^{n}$. The theorem also holds more generally for the de Sitter group \cite{tak}.

The Reeh-Schlieder theorem (theorem $4.2$ page $138$ in \cite{strweit}): Suppose $\mathcal{O}$ is an open set of the spacetime and $\mathcal{P}(\mathcal{O})$ is a $*$-algebra, {\it i.e.} the polynomial algebra of operators on $\mathcal{O}$. Then $\vert \Omega \rangle$ is a
cyclic vector for $\mathcal{P}(\mathcal{O})$, if it is a cyclic vector for $\mathcal{P}(\mathbb{R} ^{4})$.
That is, vectors of the form
$$\vert \psi \rangle \equiv \sum_{j=0}^{N} \phi(f^{(j)}_1) \cdots \phi(f^{(j)}_j)\vert \Omega \rangle ,$$
are dense in $\mathcal{H}$ with supp $f^{(k)}_j\subset \mathcal{O}$. The vacuum state $\vert \Omega \rangle$ is a cyclic vector for the field algebra $\mathcal{A}(\mathcal{O}$) corresponding to any open set $\mathcal{O}$ in Minkowski space. For the proof of this theorem, see \cite{strweit} and for the comprehensive physical consequences of this theorem, see \cite{wit}, which demonstrates that all field variables in any one region of spacetime are entangled with variables in other regions.

The Kall\'en-Lehmann spectral representation gives a general expression for the time-ordered product two-point function of an interacting quantum field theory, $G(x,x')$, as a sum of free Feynman propagators:
$$ G(x,x')=\int _{0}^{\infty }\ud \mu ^{2}\varrho (\mu ^{2})\; G_F(x,x';\mu)\, ,$$
where $\varrho (\mu ^{2})$ is the spectral density function that should be positive definite.  It is important to note that in gauge theory, the positive definite condition cannot be granted, but a spectral representation can be provided \cite{bailse}. 


\section{Bros-Fourier-Helgason transformation}  \label{apc0}
\subsection{Massive field} \label{apc}

This appendix briefly recalls the proof of equation \eqref{plan}, for precise proof see \cite{brmo2}. The Bros-Fourier-Helgason transformation, in $2$-dimensional dS spacetime for principal series, can be defined in terms of the dS plane waves as:
\b  \label{dsfou21}  \tilde f_\pm(\xi,\nu)=\int_X \ud\mu(x)\,\mbox{bv}(Hz_\pm\cdot\xi)^{-\frac{1}{2}-\ii\nu}  f(x)\equiv \langle \nu\, ;\xi_\gamma\vert f\rangle\, ,\e
where $z_+(-)$ is for $z\in \mathcal{T}^+(\mathcal{T}^-)$. The inversion of the Bros-Fourier-Helgason transformation is defined on tuboids and the integrals over the positive null cone and positive line $R^+$ as:
\b  \label{plan21} F(z)=\frac{1}{2\pi^2} \int_0^\infty \ud\nu \; \frac{\nu \tanh \pi\nu }{e^{\pm\pi\nu}\cosh \pi\nu}\int_\gamma \ud\mu_\gamma(\xi) \, \left(z\cdot \xi \right)^{-\frac{1}{2}+\ii\nu} \tilde{f}_\pm(\xi,\nu) \equiv \langle z \vert F\rangle\,.\e
The Cauchy kernel on $X^{(c)}$ is given by the following double integrals:
\b \label{infou} \frac{1}{(z'-z)^2} =-\frac{1}{2} \int_0^\infty \ud\nu \; \frac{\nu \tanh \pi\nu }{e^{\pm\pi\nu}\cosh \pi\nu}\int_\gamma \ud\mu_\gamma(\xi) \, \left(z\cdot \xi \right)^{-\frac{1}{2}+\ii\nu}\left( \xi\cdot  z' \right)^{-\frac{1}{2}-\ii\nu}\, \e
which is absolutely convergent for $(z,z')$ in tuboid $\mathcal{T}_{12}$ or $\mathcal{T}_{21}$. By replacing equation \eqref{dsfou21} in \eqref{plan21} and using equation \eqref{infou}, we obtain:
\b \label{delt2} F(z)=\frac{\pm 1}{\pi^2}\int_X \ud\mu(x)\,\frac{f(x)}{(x-z)^2} \equiv \int_X \ud\mu(x)\,\delta(x,z)f(x)\,,\e
which is a representation of the Dirac delta function in terms of Cauchy's integral on holomorphic functions. Now by inserting the resolution of the identity operator in the following relation:
$$ \langle f| g\rangle=    \langle f|\left(    \int_X \ud\mu(x)\, \vert x\rangle \langle x \vert \right)g\rangle=\langle f|\left(   \int_0^\infty d\nu\,\varrho(\nu) \int_\gamma \ud\mu_\gamma(\xi)\, \vert \nu\;;\xi_\gamma \rangle \langle\nu\,, \xi_\gamma \vert\right)g\rangle\, ,$$
and using the above integrals, equation \eqref{plan} can be proved.


\subsection{Massless field} \label{apd}

In this appendix, we calculate the proof of the equation \eqref{indsfou} in $4$-dimensions for the mcc scalar field. The Bros-Fourier-Helgason transformation, in $4$-dimensional dS spacetime, can be defined in terms of the dS plane waves of the mcc scalar field as \cite{brmo96}:
$$ \label{dsfou2} \tilde f_\pm(\xi)=\int_X \ud\mu(x)\,\mbox{bv}\left(Hz_\pm\cdot \xi \right)^{-2} f(x)\, .$$
The inversion of the Bros-Fourier-Helgason transformation is defined on tuboids and integrals over the positive null cone \cite{brmo2}:
$$ \label{plan2} F(z)=c_\pm\int_\gamma \ud\mu_\gamma(\xi) \, \left(Hz\cdot \xi \right)^{-1} \tilde{f}_\pm(\xi) \,,$$
where $c_\pm$ is constant. In this case, contrary to the principal series representation, $F$ is the solution of the mcc scalar field equation, $\left( Q_0^{dS}-2\right)F(z)=0$.

The Cauchy kernel (or analytic two-point function) on $X^{(c)}$ is given by the following integral \cite{brmo96,ta1}:
\b \label{infou2} \frac{1}{(z'-z)^2} =\frac{\ii}{8\pi^2H}\int_\gamma \ud\mu_\gamma(\xi) \, \left(z\cdot \xi \right)^{-2}\left( \xi\cdot z' \right)^{-1}\, \e
which is absolutely convergent for $(z,z')$ in tuboid $\mathcal{T}_{12}$ or $\mathcal{T}_{21}$ and then we have:
\b \label{delt4} F(z)=c_0 \int_X \ud\mu(x)\,\frac{f(x)}{(x-z)^2}\equiv \int_X \ud\mu(x)\,\delta(x,z)f(x)\,.\e
The constant $c_0$ can be precisely calculated. The validity of this equation completes the proof of the equation \eqref{indsfou} for the mcc scalar field. It is interesting to note that in this case, there is no need to add an integral over a complementary series mass parameter. With the mcc scalar field, the Cauchy integral representation viewed as a Dirac delta distribution can be obtained.

The situation is completely different for the mmc scalar field. The mmc scalar field corresponds to the representation $\Pi_{1,0}$ lying at the bottom of the discrete series representation of the dS group and the integrals over $\nu$ considered in this paper should be replaced with a sum over $n\geq 0\,;\;(Hx\cdot \xi)^n\,,\;\; (Hx\cdot \xi)^{-n-3}$ in the definition of the Bros-Fourier-Helgason transformation (in $2$-dimensional dS spacetime for discrete series see \cite{brmo2}). Although the solution $(Hx\cdot \xi)^n$ is well defined in $x\cdot \xi=0$, it diverges in the limit $x\cdot \xi =\infty$ and one cannot construct well-defined field operators for overall dS spacetime in a distributional sense. Hence one cannot define scalar field operators for discrete series representations, which means that their quantization breaks dS invariance. While a covariant quantization of the mmc scalar field can be implemented using the Krein space formalism \cite{gareta00}, the physical space of positive norm states breaks dS invariance. Viewed as a gauge potential the mmc field is a part of the quantum dS geometry, as a conformal sector of the metric \cite{ta223,tafinal}.

As a final point, the propagation of the massive particle in the principal series representation requires an integral over the representation parameter $\nu$. This may be interpreted as particle creation with different values of this parameter due to gravitational interactions. The integral or sum over a representation parameter is not required for the massless spinor field ($s=\frac{1}{2}$), the massless vector field ($s=1$), and the massless vector-spinor field ($s=\frac{3}{2}$), which can be constructed from the mcc scalar field and propagated on the dS light cone \cite{ta1,gagarota}. For these fields, we have the creation of massless particles, which only propagate on the dS light cone. However, we need a sum over the parameter $n$ in the discrete series representation of the massless tensor field ($s = 2$) since it is also written in terms of the mmc scalar field \cite{ta1}. In this case, similar to the mmc scalar field, for which we need to implement Krein space quantization.

\end{appendix}

\end{document}

\bibitem{bida} N.D. Birrell and P.C.W. Davies, Cambridge Univ. Press (1982) {\it Quantum Fields in Curved Space}.

\red{\mbox{bv}}